\documentclass[reprint, 10pt, a4paper, prx, floatfix, longbibliography]{revtex4-2}
\usepackage{amsfonts}
\usepackage{graphicx}
\usepackage{amsmath}
\usepackage[hidelinks]{hyperref}
\begin{document}

\title{Efficiency and Physical Limitations of Adiabatic  Direct Energy Conversion in Axisymmetric Fields}
\date{\today}
\author{J.-M. Rax}
\email[Electronic mail: ]{jean-marcel.rax@universite-paris-saclay.fr}
\affiliation{Andlinger Center for Energy and Environment, Princeton University, Princeton University, Princeton NJ 08544, USA}
\affiliation{IJCLab, Universit\'{e} de Paris-Saclay, 91405 Orsay, France}
\author{E.~J. Kolmes}
\affiliation{Department of Astrophysical Sciences, Princeton University, Princeton NJ 08544, USA}
\author{N.~J. Fisch}
\affiliation{Department of Astrophysical Sciences, Princeton University, Princeton NJ 08544, USA}

\begin{abstract}
	We describe and analyze a new class of direct energy conversion schemes based on the adiabatic magnetic drift of charged particles in axisymmetric magnetic fields. The efficiency of conversion as well as the geometrical and dynamical limitations of the recoverable power are calculated. The geometries of these axisymmetric field configurations are suited for direct energy conversion in radiating advanced aneutronic reactors and in advanced divertors of deuterium-tritium tokamak reactors. The E cross B configurations considered here do not suffer from the classical drawbacks and limitations of thermionic and magnetohydrodynamic high temperature direct energy conversion devices. 
\end{abstract}

\maketitle

\section{Introduction}

Part of the internal energy content of a thermodynamical system can be
extracted to produce useful work . Under reversible extraction conditions
the allowed maximum extracted work is given by the free energy difference
between the initial and final states \cite{Guggenheim, Callen}. The \textit{free energy} \textit{%
content} of a system is always lower than its total \textit{internal energy
content}. Their difference is proportional to the temperature and the
entropy which are always positive. Moreover, entropy production due to
irreversible operations decreases the fraction of the internal energy
that can be converted into useful work.

\begin{figure*}
	\includegraphics[width=\linewidth]{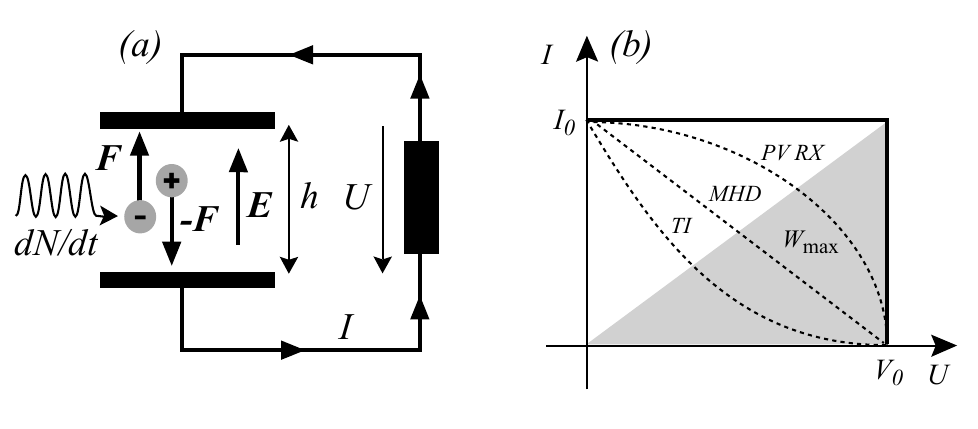}
	\caption{(a) Basic layout of a DC electric power DEC generator, (b) ideal and non-ideal current-voltage characteristics (corresponding to the solid and dashed curves, respectively).} 
	\label{fig:fig1}
\end{figure*}

Energy conversion underlies many of the technologies that have been proposed
to decarbonize the world's energy infrastructure. In addition to economic
and environmental considerations, the efficiency of free energy extraction
is one of the most important factors determining the relative merits of different
technologies \cite{Sorensen}.

Thermodynamical energy conversion systems are usually based on simple non
equilibrium states displaying gradients of intensive variables. The most common thermodymamical non equilibrium states used as free energy sources in classical conversion devices involve (\textit{i})
pressure, (\textit{ii}) temperature and (\textit{iii}) chemical potential
gradients \cite{Decher}. Pressure gradients
can be relaxed in turbines to produce mechanical work with a high conversion
efficiency. Steady state free energy extraction associated with a simple temperature differential is limited by the Carnot's reversible efficiency at
zero power and by the Curzon-Alborn-Novikov-Chambadal endoreversible
efficiency at maximum power \cite{Novikov1957, Chambadal, Curzon1975}. Chemical potential differences can be
efficiently converted in electrochemical devices but are, unfortunately,
usually relaxed through open air combustion to sustain a temperature
gradient in combustion driven systems.

Besides classical thermal conversion schemes, direct energy conversion (DEC)
schemes such as magnetohydrodynamic (MHD) generators, thermoionic diodes
(TI), photovoltaic (PV) and redox (RX) cells of the hydrogen type have been
recognized to offer the potential of significant conversion efficiency as
they avoid the inefficient steps requiring the cooling/heating of a
compressed/expanded gas \cite{Rax2015Book, Sutton, DecherDEC}. The basic principles of a DC electric
power DEC generator is illustrated in Fig.~\ref{fig:fig1}-(a). Two steps are required: (%
\textit{i}) free charge generation at a rate $dN/dt$ followed by (\textit{ii%
}) charge separation with a force $\pm F$ on positive and negative
particles. The resulting flow of current must be oriented such
that it provides power to an electric field $E$ sustained between two
electrodes separated by a gap width $h$.

A device's current-voltage ($I$-$U$) characteristic describes how its efficiency changes as its power throughput increases. 
Under ideal operations, the current-voltage ``square'' characteristic of
this ideal DEC generator is illustrated in Fig.~\ref{fig:fig1}-(b). The short circuit
current $I_{0}$ is given by
\begin{equation}
I_{0}=2q \frac{dN}{dt}
\end{equation}
where $q$ is the single particle charge. The open circuit voltage
\begin{equation}
V_{0}=\frac{Fh}{q}
\end{equation}
is reached when the electric force $\pm qE=\pm qU/h$ balances the separating
force $\pm F$.

When non ideal processes are taken into account the ``square'' ideal
characteristic becomes either a curve of the concave (TI) or convex (PV and
RX) type, or simply a resistive straight line (MHD), depicted by the three
dotted curves in Fig. 1-(b) \cite{Rax2015Book}. The maximum power $W_{\max }$ that can be
delivered to the external load from an ideal generator is 
\begin{equation}
W_{\text{max} }=I_{0}U_{0}/2=Fh\left( dN/dt\right) \text{.}
\end{equation}
The force $F$ used in DEC devices is typically of a statistical nature: thermodynamical
forces associated with the gradient of an intensive variable such as pressure,
temperature or chemical potential. In the new class of plasma DEC presented
and analyzed here the force $F$ is the centrifugal force due to the
curvature of the magnetic field lines and the diamagnetic force due to the
gradient of the magnetic field strength. These are in fact thermal forces
because their effects are proportional to the kinetic/thermal energy of the
particles. This kinetic/thermal energy can be transformed into DC electric
power through the design of a dedicated field configuration.

In TI diodes $F$ is associated with the pressure/temperature gradient of the
free electrons emitted by the hot cathode. Ions remain bounded in the
cathode lattice and sustain a phonon gas rather than flowing. In MHD
generators $F$ is the friction force of the flowing neutral part of the
weakly ionized plasma. In PV generators $F$ is a chemical potential gradient
resulting from the chemical potential difference between the two sides of
the PV junction.


The type of DEC generator illustrated in Fig. 1-(a) is well suited when the
separating force $F$ acts in opposite directions on electrons and ions. If
the force $F$ is insensitive to the sign of the charges we can either: (%
\textit{i}) immobilize one type of particle or (\textit{ii}) use an E cross
B configuration where the $\mathbf{F}\times \mathbf{B}/qB^{2}$ drift
velocity separates the charges. An ideal E cross B configuration is
illustrated in Fig. 2. For a small voltage drop $U$ and a steady state
population of $2N$ charged particles in the electrode gap, the ideal power $%
W_{0}$ of the generator is the product of the drift velocity $F/qB$ times
the electric force $qE$: 
\begin{equation}
W_{0}=2NFE/B\text{.}
\end{equation}

\begin{figure}
	\includegraphics[width=\linewidth]{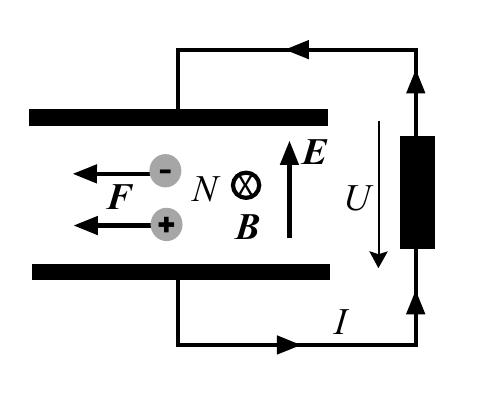}
	\caption{An illustration of an $\mathbf{E} \times \mathbf{B}$-type configuration. A force $\mathbf{F}$ acts similarly on both positive and negative charged particles, so the resutling $\mathbf{F} \times \mathbf{B}$ drifts separate the charges. }
	\label{fig:fig2}
\end{figure}

The past decades witnessed the massive development of PV and RX cell DEC
generators which are now produced on a fully developed industrial scale.
High temperature TI and MHD DEC generators performances remain below the
requirement to envision an industrial scale development. Despite the lack of
industrial scale achievements, the continuous interest for TI and MHD stems
from the fact that they operate at high temperatures: (\textit{i}) for a
given amount of energy, high temperature heat offers the potential of a far
better conversion than low temperature heat, (\textit{ii}) for the same
power high temperature MHD and TI systems occupy a smaller footprint than
classical systems.

\begin{figure*}
	\includegraphics[width=\linewidth]{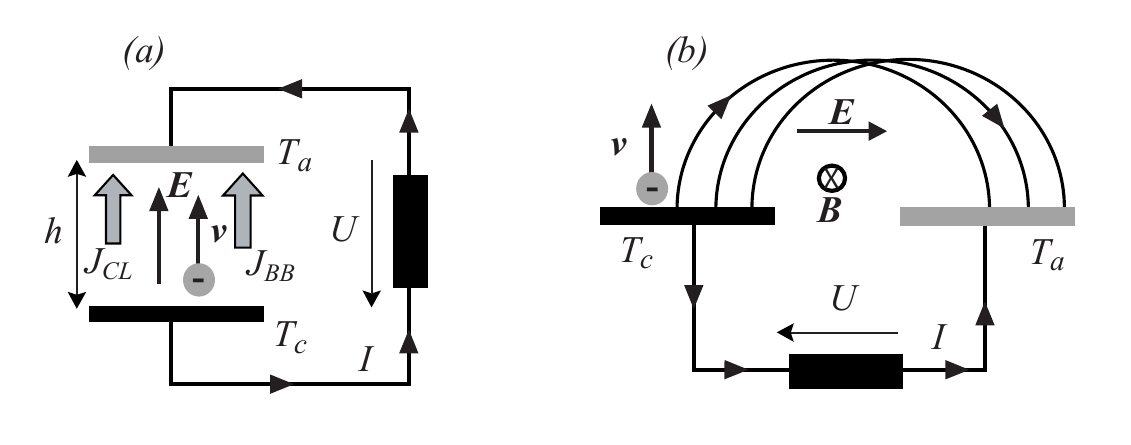}
	\caption{Two thermionic diode configurations: (a) the basic configuration, in which thermionic emission ejects electrons parallel to an electric field, and (b) a modified configuration, in which a magnetic field is used to avoid the Child-Langmuir and blackbody limitations on the attainable current. }
\end{figure*}

Very high temperature heat is produced in thermonuclear reactors, (\textit{i}%
) in the form of a high temperature plasma flow at the level of the divertor
in $^{2}$D/$^{3}$T tokamak reactors, or (\textit{ii}) in the form of high
intensity short wavelength radiation in advanced neutronless $^{1}$P/$^{11}$%
B reactors. An efficient high temperature DEC scheme would be very
beneficial for $^{2}$D/$^{3}$T and $^{1}$P/$^{11}$B fusion schemes. Several
processes have been put forward to achieve direct conversion of thermonuclear energy \cite{Perkins1988}, such as cusp configurations \cite{Takeno2019}, traveling waves \cite{Momota1999} and advanced electrostatic configurations \cite{Yoshikawa1988, Volosov2005} or electrostatic and magnetostatic configurations of the E cross B type \cite{Timofeev1978}. In this work we analyze a class of high temperature E cross B type DEC scheme free from the usual drawbacks of MHD and TI devices. The drawbacks of TI and MHD high temperature DEC devices have been known for a long time. To mention a few: the occurrence of space charge limited flow in vacuum TI diodes and the erosion of the edge electrodes in MHD generators put severe restrictions on the
efficiency of such generators.

For example, let us consider a thermionic diode illustrated in Fig. 3-(a). A
high-temperature source sustains a temperature difference between a hot
cathode ($T_{c}$) and a cold anode ($T_{a}\ll T_{c}$). Thermionic emission,
described by Richardson-Dushman's law \cite{Dushman1923}, takes place at the inner
surface of the cathode and electrons, with mass $m$, work against an
electric field $E$ during their transit from the cathode toward the anode.

Under optimal conditions the voltage $U$ of such an electric generator is
given by the relation $mv^{2}=2qU$ where $v\sim \sqrt{k_{B}T_{c}/m}$ is the
average emission velocity. However, the heat driven current $I$ is limited by
Child-Langmuir's law, limiting the current density to a value $J_{CL}\left[ 
\text{A/m}^{2}\right] $given by 
\begin{equation}
J_{CL}=2\varepsilon _{0}mv^{3}/9qh^{2}\text{,}
\end{equation}
where $h$ is the anode-cathode gap width. Thus, to extract significant
power, an impractically tiny gap is needed. Moreover, beside the electron flux
described by $J_{CL}$, the black-body flux of photons $J_{BB}\left[ \text{W/m%
}^{2}\right] $ provides a thermal short circuit dramatically lowering the
conversion efficiency when $h$ is small, 
\begin{equation}
J_{BB}=\pi ^{2}k_{B}^{4}T_{c}^{4}/15\hbar ^{3}c^{2}\text{, }
\end{equation}
where $\hbar $ is Planck's constant. To avoid these drawbacks an E cross B
configuration, illustrated in Fig. 3-(b), has been proposed \cite{Hatsopoulos}. The
current across the magnetic $B$ field is no longer limited to $J_{CL}$ and $%
J_{BB}$ no longer heats up the cold anode, but the ballistic coupling between
the cathode and the anode turn out to be inefficient because of the
dispersion in the velocities of the electrons. Other mitigations of the TI drawbacks, such as the use of plasma TI diode rather than vacuum TI diodes \cite{Balsht1978}, have been considered but none of them have made it possible to achieve the expected high conversion efficiency.

\begin{figure*}
	\includegraphics[width=\linewidth]{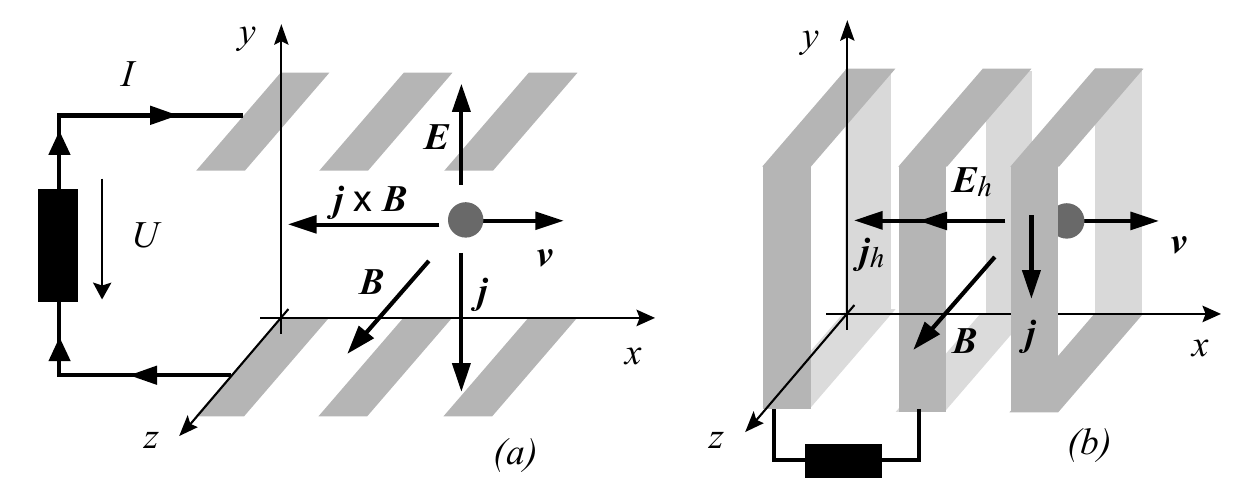}
	\caption{Two MHD generator configurations: (a) the Faraday configuration and (b) the Hall configuration.} 
\end{figure*}

Nevertheless, E cross B configurations have proven their usefulness in MHD
generator designs used to convert the free enthalpy of a hot weakly
ionized plasma flow into DC electric power. This E cross B configuration is
illustrated in Fig. 4-(a). A weakly ionized plasma flows, with a velocity $\mathbf{v}$ along $x$, across a magnetic field $\mathbf{B}$ directed along
the $z$ axis \cite{Nedospasov1977}. Electrons and ions set up a current $\mathbf{j}$ under
the influence of the $\mathbf{v} \times \mathbf{B}$ force. This current
provides power to the electric field $\mathbf{E}$ along the $y$ axis. Energy
conservation is ensured as the $\mathbf{j} \times \mathbf{B}$ force slows
down the flow along the $x$ axis. The short circuiting of the $y$ currents
provides another E cross B configuration beside the segmented Faraday
configuration illustrated in Fig. 4-(a): the Hall configuration illustrated
in Fig. 4-(b). The basic physical principles and main limitations of Faraday
and Hall MHD DEC generators can be found in classical textbooks Ref.~\cite{SuttonGenerators, Rosa}.
Despite the simplicity and effectiveness of the physical principles put at
work in Faraday and Hall generators, the management of a hot weakly ionized
collisional plasma flow has proven to be difficult and MHD generators did
not find their way to industrial development up to now.

The use of $\mathbf{E}\times \mathbf{B}$ configurations aimed at DEC is not
restricted to advanced TI and classical MHD generators illustrated in
Figures~3 and 4. We will describe and analyze in this paper another $%
\mathbf{E}\times \mathbf{B}$ configuration where the thermal energy of
charged particles is converted to a DC electromotive force in very
particular types of inhomogeneous magnetic and electric fields.

This new configuration does not suffer from the major drawbacks of the TI
scheme and the MHD scheme. The Child-Langmuir law limitation does not apply
and electrode erosion is minimized as the charged particles strike the
electrodes at low energy. This configuration also has its own advantages: (%
\textit{i}) the energy extraction rate is exponential with respect to time
and (\textit{ii}) the closed field line topology minimizes plasma losses.
Besides the topology, the geometry is particularly pertinent for high
temperature conversion in fusion reactors. 
The proposed configuration can be understood as a way of converting plasma kinetic energy into electricity. 
It can also be understood as a technique for capturing radiation, if that radiation is used to ionize neutrals and the energy is captured from the resulting charged particles. 

This paper is organized as follows. In the next section we describe the way
to arrange coils and electrode plates in poloidal and toroidal axisymmetric
configurations in order to extract the thermal energy of a plasma. We
analyze the energy exchange between charged particles and the electric field
which provide the electromotive force of the generator in these new $\mathbf{%
E}\times \mathbf{B}$ configurations in Sections III and IV. We calculate the
efficiency and the irreducible physical limitations on the power delivered
by toroidal DEC generator in Sections V and VI. We do not consider the
limitations associated with the stress on the material in a high temperature
environment. This paper is instead devoted to an analysis of the physical
principles. The adaptation of these DEC scheme to $^{2}$D/$^{3}$T tokamak
reactors and $^{1}$P/$^{11}$B advanced reactors is briefly considered in
Section VII. The last section summarizes our new results and gives our
conclusions.

\section{E cross B cooling in poloidal and toroidal configurations}

The design of a hot plasma DEC device requires the identification of a
structure such that the motion of the charged particles is slowed down by an
electric field sustained between two electrodes. As a result of global
energy conservation this E cross B electric cooling of a hot plasma results
in the sustainment of an electromotive force when the electrode circuit is
closed on an external load.

\begin{figure*}
	\includegraphics[width=\linewidth]{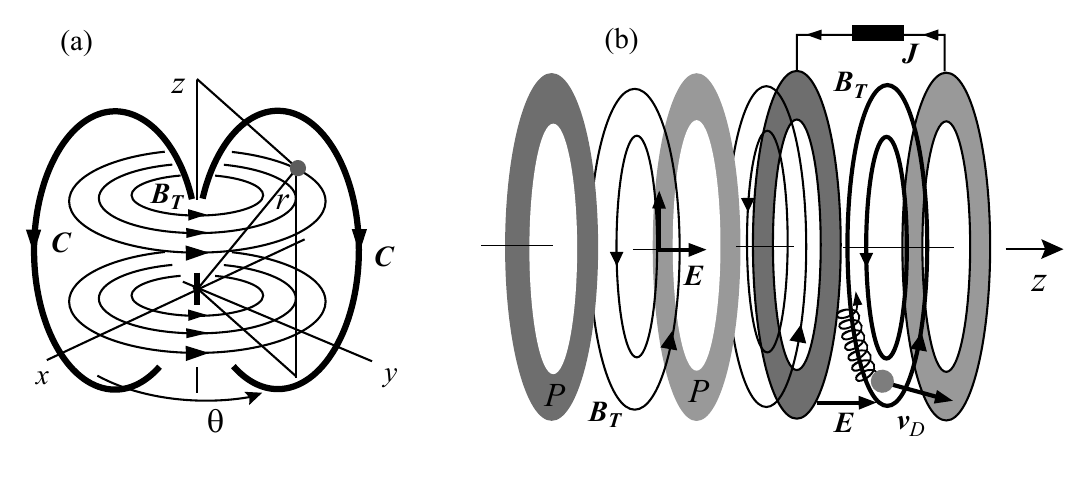}
	\caption{A toroidal field configuration for DEC; (a) shows the positioning of the coils $C$ used to generate the magnetic field and (b) shows the current-collecting electrode plates $P$ that generate the electric field. }
\end{figure*}

\begin{figure*}
	\includegraphics[width=\linewidth]{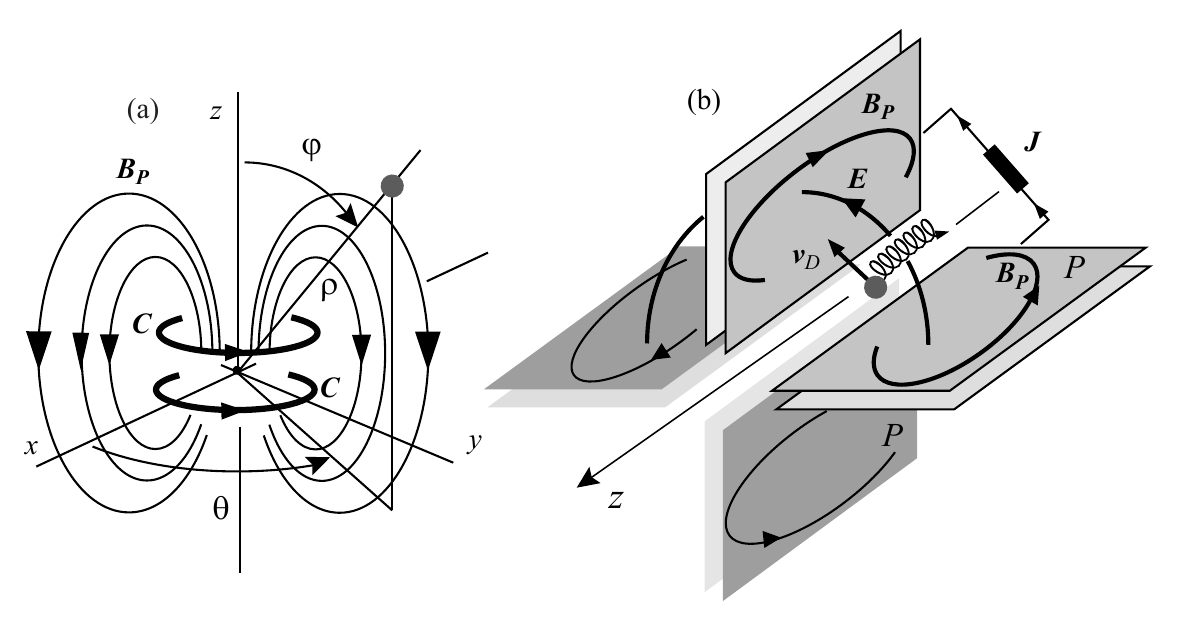}
	\caption{A poloidal field configuration for DEC; again, (a) shows the positioning of the magnetic field coils $C$ and (b) shows the electrode plates $P$. }
\end{figure*}

The magnetic drift velocity is proportional to $\mathbf{B}\times \mathbf{%
\nabla }B$ and to the thermal enegy content of the plasma particles. The
sign of the resulting power transfer between the hot plasma and the electric field is
controlled by the sign of the dot product between the drift velocity and the
electric field. The condition for plasma cooling and DC power generation is 
\begin{equation}
\mathbf{E}\cdot \mathbf{B}\times \mathbf{\nabla }B<0 . 
\end{equation}
This can be adjusted in axisymmetric magnetic field configurations
through the electrode plates' $P$ positions, shapes and polarizations, so
plasma cooling and electric power generation can be envisioned and will be
studied in the next sections. In the axisymmetric magnetic field
configurations depicted on Fig. 5-(a) and Fig. 6-(a) the charged particle
motion is the drift motion in an inhomogeneous magnetic field and the
electric field is sustained between electrode plates $P$ collecting the
drift current $\mathbf{J}$. The azimuthal angle around the Cartesian axis $%
z $ is denoted by $\theta $ ($0\leq \theta <2\pi $) and the unit vector $\mathbf{%
e}_{\theta }$ corresponds to this azimuthal direction. Any axisymmetric
magnetic field $\mathbf{B}$ can be represented as the sum of a poloidal plus
a toroidal field: $\mathbf{B}=B_{T}\mathbf{e}_{\theta }+\mathbf{\nabla }%
\times \left( A_{P}\mathbf{e}_{\theta }\right) $ where the first term on the
right hand side is the toroidal component and the second is the poloidal
component. Thus we consider two types of structures aimed at extracting the
free energy of a hot plasma and converting it in DC electric power:
toroidal and poloidal configurations illustrated respectively in Fig. 5 and
Fig. 6. For toroidal generators the electric field is axial, and for
poloidal generators the electric field is azimuthal.

In Fig. 5-(b) the $z=$ constant ring-shaped conducting electrodes $P$ are
used to collect the drift current $\mathbf{J}$ in the axial direction. In
Fig.6-(b) the $\theta =$ constant plane electrodes $P$ are used to collect
the drift current $\mathbf{J}$ in the azimuthal direction. With these
orientations of the electric and magnetic fields, the adiabatic magnetic
drift velocity of the hot charged particles can be directed against the
electric field's force to extract thermal energy and cool down the plasma.
In both cases this drift current $\mathbf{J}$ provides DC power $\mathbf{J}%
\cdot \mathbf{E}$ to the external load.

A complete analysis must also take into account the electric $\mathbf{E}%
\times \mathbf{B}/B^{2}$ drift. We will see that the impact of this drift
is to facilitate energy conversion at low values of $E$ and inhibit it at
larger values.

An axisymmetric vacuum magnetic field, $\mathbf{\nabla }\times \mathbf{B}=%
\mathbf{0}$, $\mathbf{\nabla }\cdot \mathbf{B}=0$, $\partial \mathbf{B}%
/\partial \theta =\mathbf{0}$, either poloidal or toroidal, is locally
represented on the Frenet-Serret basis associated with the magnetic field
line as $\mathbf{B}=B\mathbf{b}$. The gradient $\mathbf{\nabla }B$
includes two terms

\begin{equation}
\mathbf{\nabla }B=\frac{\partial B}{\partial s}\mathbf{b}+\frac{B}{R}\mathbf{%
n}  \label{FS}
\end{equation}
where $R$ is the curvature radius of the field line, $s$ is the curvilinear
abscissa along the field line and $\left( \mathbf{b=B}/B,\mathbf{n}=\partial 
\mathbf{b}/\partial s\right) $ are the tangent and normal unit vectors to
the field line: the Frenet-Serret moving frame without torsion. The $%
\mathbf{b}$ component in (\ref{FS}) generates the diamagnetic force along
the field lines and the $\mathbf{n}$ component the drift velocity.

In Section III, toroidal magnetic fields will be conveniently described with
a cylindrical set of coordinates $\left[ r,\theta ,z\right] $ rather than
with the Cartesian one $\left[ x,y,z\right] $ ($x=r\cos \theta $, $y=r\sin
\theta $). The toroidal magnetic field, displayed in Fig. 5-(b), is assumed
to be without ripple despite the finite number of coils $C$ 
\begin{eqnarray}
\mathbf{B} &=&B\mathbf{e}_{\theta }=B_{0}\frac{r_{0}}{r}\mathbf{e}_{\theta }%
\text{,}  \label{fs23} \\
\mathbf{E} &=&-\mathbf{\nabla }\phi =-E\mathbf{e}_{z}\text{,}  \label{fs32}
\end{eqnarray}
where $r_{0}$, $E$, and $B_{0}$ are positive constants. Here we have used
Amp\'{e}re's theorem, $Br=B_{0}r_{0}$ and introduced the electric potential $%
\phi $. The gradient of the magnetic field strength is directed along the
radial direction 
\begin{equation}
\mathbf{\nabla }B=-\frac{B}{r}\mathbf{e}_{r} \, .  \label{fs24}
\end{equation}
For a purely toroidal field $\mathbf{n=-e}_{r}$, $\partial B/\partial s=0$
and $R=r$. The poloidal magnetic field configuration illustrated in Fig. 6
and analyzed in Section IV is usually described with spherical
coordinates $\left[ \rho ,\varphi ,\theta \right] $ rather than
cylindrical coordinates $\left[ r,\theta ,z\right] $ ($r=\rho \sin \varphi $, $%
z=\rho \cos \varphi $). The azimuthal electric field depicted in Fig. 6-(b)
can be approximated by 
\begin{equation}
\mathbf{E}=- E\mathbf{e}_{\theta }=-\mathbf{\nabla }\phi \, . \label{epol}
\end{equation}
The full expression for the field from the electrodes in the figure would include some additional terms, but the simple form given in Eq.~(\ref{epol}) is sufficient to show the essential behavior of the energy transfer mechanism. 
For the poloidal case, we will assume (\textit{i}) that the source of the
thermal plasma is restricted to the region near the equatorial plane $%
\varphi \sim \pi /2$ in between each pair of plates and (\textit{ii}) that
the field geometry and the conducting electrode plates $P$ are designed
such that the capture of the positive and negative charges, above and below
this equatorial plane, takes place at a small angle $\varphi $. It is
convenient to consider the Frenet-Serret representation Eq. (\ref{FS}) in which the
radius of curvature of the field lines is denoted by $R$ and the gradient scale length $%
L$ along the field line is defined as $B/L=\partial B/\partial s$. Along a
given field line both the radius $R$ and the length $L$ are functions of $s$
and we will describe the gradient of the magnetic field in the drift region,
above and below the equatorial plane, along and across the field lines
between the electrodes, with the model
\begin{equation}
\mathbf{\nabla }B=\frac{B}{L}\mathbf{b}+\frac{B}{R}\mathbf{n}  \label{appr}
\end{equation}
where rather than the exact functions $L\left( s\right) $ and $R\left(
s\right)$ we consider the average over $s$ of $L\left( s\right) $ and $%
R\left( s\right)$ in the region explored by the charged particles in
between the equatorial plane and the ultimate capture by the electrodes $P$
at a small angle $\varphi $. 
This approximation makes it simpler to show the dependence of the conversion process on $R$ and $L$ in cases where both are important. 

\section{Adiabatic thermal energy conversion in toroidal field}

\begin{figure}
	\includegraphics[width=\linewidth]{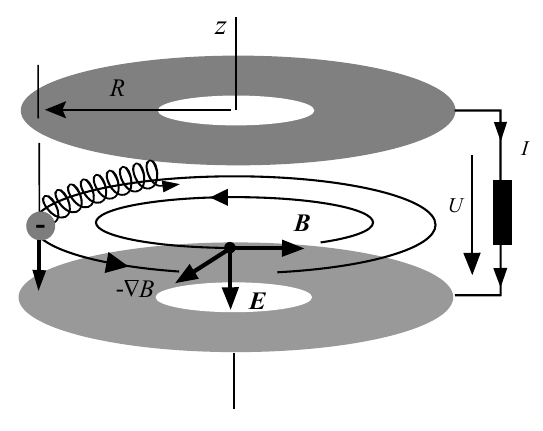}
	\caption{Schematic of the fields associated with a toroidal field, including the magnetic field-strength gradient and the positions of the current-collecting electrodes.}
\end{figure}

Consider a cylindrical set of coordinates $\left[ R,\theta ,z\right] $
associated with the cylindrical basis $\left[ \mathbf{e}_{r},\mathbf{e}%
_{\theta },\mathbf{e}_{z}\right] $. The position of a particle with charge $%
q $ and mass $m$ is given by $R\mathbf{e}_{r}\left( \theta \right) +z\mathbf{%
e}_{z}$. The field geometry between two electrodes is illustrated in Fig. 7.
We will calculate the thermal energy extraction resulting from the drift-driven electric cooling. The E cross B configuration Fig. 7 is described by Eqs. (\ref{fs23},\ref{fs32}) where we take $r=R$.

The gradient of the azimuthal magnetic field strength is radial and is described by Eq. (\ref{fs24}) 
\begin{equation}
\mathbf{\nabla }B=\mathbf{-}B_{0}\frac{R_{0}}{R^{2}}\mathbf{e}_{r}=\mathbf{-}%
\frac{B}{R}\mathbf{e}_{r}\text{.}
\end{equation}
The motion of a charged particle in such a toroidal field configuration is
a combination of a translation $v_{\parallel }$ along the magnetic field
lines, cyclotron rotation $v_{\perp }$ around the field lines and drift $%
\mathbf{v}_{D}$ across the magnetic field lines. The total velocity $\mathbf{%
v}$ is thus given by 
\begin{equation}
\mathbf{v}=v_{\parallel }\frac{\mathbf{B}}{B}+\mathbf{v}_{D}+v_{\perp }\cos
\left( \omega _{c}t\right) \mathbf{e}_{r}+v_{\perp }\sin \left( \omega
_{c}t\right) \mathbf{e}_{z}
\end{equation}
where $\omega _{c}=qB/m$ is the cyclotron frequency. The drift velocity $%
\mathbf{v}_{D}$ is a combination of the magnetic and electric drifts 
\begin{equation}
\mathbf{v}_{D}=\frac{2\varepsilon _{\parallel }+\varepsilon _{\perp }}{qB^{3}%
}\mathbf{B}\times \mathbf{\nabla }B+\frac{\mathbf{E}\times \mathbf{B}}{B^{2}}
\label{adia1}
\end{equation}
where we have introduced $\varepsilon _{\parallel }$ = $mv_{\parallel
}^{2}/2 $, the kinetic energy along the field lines, and $\varepsilon
_{\perp }$ = $mv_{\perp }^{2}/2$ the cyclotron kinetic energy around the
field lines. The axial and radial drift equations are given by

\begin{eqnarray}
\frac{dz}{dt} &=&\frac{2\varepsilon _{\parallel }+\varepsilon _{\perp }}{%
qB^{3}}\mathbf{B}\times \mathbf{\nabla }B\cdot \mathbf{e}_{z}  \label{drif1}
\\
\frac{dR}{dt} &=&\frac{\mathbf{E}\times \mathbf{B}}{B^{2}}\cdot \mathbf{e}%
_{r}  \label{drif2}
\end{eqnarray}
The power transfer between the thermal energy and the electric energy is
given by the dot product 
\begin{equation}
q\mathbf{E}\cdot \mathbf{v}_{D}=\frac{2\varepsilon _{\parallel }+\varepsilon
_{\perp }}{B^{3}}\mathbf{B}\times \mathbf{\nabla }B\cdot \mathbf{E}=-\frac{%
2\varepsilon _{\parallel }+\varepsilon _{\perp }}{\tau }  \label{trener}
\end{equation}
where we have introduced the secular time scale $\tau >0$ defined as 
\begin{equation}
\tau \doteq \left| \frac{B^{3}}{\mathbf{E}\times \mathbf{B\cdot \nabla }B}%
\right| =\frac{B_{0}R_{0}}{E}\text{.}  \label{to}
\end{equation}
The potential energy $q\phi =qEz$ increases in time at the expense of the
thermal energy $\varepsilon _{\parallel }+\varepsilon _{\perp }$ so that the
sum of the kinetic plus potential energy $\varepsilon $ remains constant.
Two invariants can be identified: the magnetic moment $\mu $ is an adiabatic
invariant and the energy $\varepsilon $ is a Noether invariant. The cyclotron
and total energies can be written as 
\begin{eqnarray}
\varepsilon _{\perp }\left( R\right) &=&\mu B\left( R\right) \text{,} \\
\varepsilon \left( z,R\right) &=&\varepsilon _{\parallel }+\varepsilon
_{\perp }+q\phi \left( z\right) \text{.}  \label{energy}
\end{eqnarray}
We have used the adiabatic ordering and neglected the small drift kinetic
energy. The case of a strong electric field, where we take this drift energy
into account, is considered in Section VI. Thus as $d\mu /dt=0$ and $%
d\varepsilon /dt=0$, the adiabatic evolution of the energy is described by 
\begin{eqnarray}
\frac{d\varepsilon _{\perp }}{dt} &=&\left[ \left( \mathbf{v}%
_{D}+v_{\parallel }\frac{\mathbf{B}}{B}\right) \cdot \mathbf{\nabla }\right]
\mu B\text{,}  \label{ener2} \\
\frac{d\varepsilon _{\parallel }}{dt} &=&-\left[ \left( \mathbf{v}%
_{D}+v_{\parallel }\frac{\mathbf{B}}{B}\right) \cdot \mathbf{\nabla }\right]
\left( q\phi +\mu B\right) \text{.}  \label{ener3}
\end{eqnarray}
We consider here the slow evolution averaged over the fast cyclotron motion
and we have eliminated $\varepsilon _{\perp }$ to set up Eq. (\ref{ener3})
as $d\varepsilon /dt=0$. The secular velocity operator involved in Eqs. (\ref
{ener2},\ref{ener3}) is given by 
\begin{equation}
\left( \mathbf{v}_{D}+v_{\parallel }\frac{\mathbf{B}}{B}\right) \cdot 
\mathbf{\nabla }=\frac{2\varepsilon _{\parallel }+\varepsilon _{\perp }}{qRB}%
\frac{\partial }{\partial z}+\frac{E}{B}\frac{\partial }{\partial R}\text{,}
\label{adia2}
\end{equation}
where we note that the $v_{||}$ term drops out due to axisymmetry. We
substitute this secular velocity given in Eq. (\ref{adia2}) into Eq. (\ref
{ener2},\ref{ener3}) to get the thermal energy extraction dynamical
equations 
\begin{eqnarray}
\frac{d\varepsilon _{\parallel }}{dt} &=&-2\frac{\varepsilon _{\parallel }}{%
\tau }  \label{e1} \\
\frac{d\varepsilon _{\perp }}{dt} &=&-\frac{\varepsilon _{\perp }}{\tau } \, . 
\label{e2}
\end{eqnarray}
We recover the energy balance from Eq. (\ref{trener}).

During the transit of one charged particle toward the electrode, its thermal
energy decreases at an exponential rate with respect to time. This behavior
provides an efficient way to directly extract the thermal energy. The
physics behind this exponential extraction of the thermal energy 
\begin{equation}
\varepsilon _{\parallel }+\varepsilon _{\perp }=\varepsilon _{\parallel
0}\exp \left( -\frac{2t}{\tau }\right) +\varepsilon _{\perp 0}\exp \left( -%
\frac{t}{\tau }\right)  \label{cool}
\end{equation}
can be described as follows.

The electric drift $E/B$ is radial and pushes particles toward the lower-$B$
region where $\varepsilon _{\perp }$ = $\mu B$ is converted into $%
\varepsilon _{\parallel }$ in order to ensure the adiabatic invariance of $%
\mu $. At the very same time the magnetic drift is axial along $z$ and
pushes particles toward high potential $q\phi $ regions where $\varepsilon
_{\parallel }+\varepsilon _{\perp }$ decreases in order to ensure the
invariance of $\varepsilon =\varepsilon _{\parallel }$ + $\varepsilon
_{\perp }$ + $q\phi $. This thermal energy extraction is illustrated in
Figure~8.

The drift equations given by Eqs.~(\ref{drif1}) and (\ref{drif2}) can be
integrated to give

\begin{equation}
z=z_{0}+\frac{\varepsilon _{\parallel 0}}{qE}\left[ 1-\exp \left( -\frac{2t}{%
\tau }\right) \right] +\frac{\varepsilon _{\perp 0}}{qE}\left[ 1-\exp \left(
-\frac{t}{\tau }\right) \right]  \label{zz}
\end{equation}
and 
\begin{equation}
R=R_{0}\exp \left( \frac{t}{\tau }\right) \text{.}
\end{equation}
These two expressions determine the dimensions of a device needed to access
a given extraction efficiency. These relations describe an expansion of the
hot plasma and they set limitations on the full thermal energy extraction as
the device must display a finite size and footprint. These geometrical
limitations will be analyzed in Section V.

\section{Adiabatic thermal energy conversion in poloidal fields}

Before addressing the geometrical and dynamical limitations of
the conversion efficiency of the process described by Eq.~(\ref{cool}), we
explore in this section the main difference between (\textit{i}) plasma
cooling in a toroidal field and (\textit{ii}) plasma cooling in a poloidal
field. Using the model poloidal field described by Eqs. (\ref{epol},\ref{appr}), the
set of relations Eqs. (\ref{e1},\ref{e2}) describing adiabatic slowing down
becomes 
\begin{eqnarray}
\frac{d\varepsilon _{\parallel }}{dt} &=&-2\frac{\varepsilon _{\parallel }}{%
\tau }-v_{\parallel }\frac{\varepsilon _{\perp }}{L}\text{,}  \label{cool23}
\\
\frac{d\varepsilon _{\perp }}{dt} &=&-\frac{\varepsilon _{\perp }}{\tau }%
+v_{\parallel }\frac{\varepsilon _{\perp }}{L}\text{.}
\end{eqnarray}
We recover the general energy balance from Eq.~(\ref{trener}), which is in fact
independent of the configuration, poloidal, toroidal or mixed, although in general 
$\tau$ may become a function of $s$. Compared to Eqs. (\ref{e1},\ref{e2}),
there is an additional term due to the diamagnetic mirror force
redistributing the energy between the parallel and cyclotron degree of
freedom to ensure $\mu $ adiabatic invariance. It is to be noted that, as
opposed to the previous toroidal case where the energy relations Eqs. (\ref
{e1},\ref{e2}) are exact in a perfect toroidal field, the poloidal case here
is only analyzed on the basis of the approximate model poloidal field Eq. (%
\ref{appr}). The aim of this section is to identify the impact of a gradient
along the field lines.

The additional $L$ term is treated as a small perturbation so that the zero
order solutions are just the solutions of Eqs. (\ref{e1},\ref{e2}) which
fulfills the conservation relation 
\begin{equation}
\varepsilon _{\perp }/\sqrt{\varepsilon _{\parallel }}=\varepsilon _{\perp
0}/\sqrt{\varepsilon _{\parallel 0}}\text{.}
\end{equation}
We introduce the characteristic time 
\[
\tau _{0}=\frac{L}{\varepsilon _{\perp 0}}\sqrt{2m\varepsilon _{\parallel 0}}
\]
and assume $\tau <\tau _{0}$. Within the framework of this perturbative
expansion the cooling of the parallel energy Eq. (\ref{cool23}) becomes 
\begin{equation}
\frac{d\varepsilon _{\parallel }}{dt}=-2\frac{\varepsilon _{\parallel }}{%
\tau }\pm 2\frac{\varepsilon _{\parallel }}{\tau _{0}}\text{.}
\end{equation}
The analysis of the cooling/conversion process takes place in the $\left(
\varepsilon _{\parallel },\varepsilon _{\perp }\right) $ energy space. In
the toroidal case the \textit{cooling trajectories} in energy space,
Eqs. (\ref{e1},\ref{e2}), are all restricted to parabolic curves in this
space 
\begin{equation}
\frac{d\varepsilon _{\parallel }}{\varepsilon _{\parallel }}=2\frac{%
d\varepsilon _{\perp }}{\varepsilon _{\perp }}\text{.}
\end{equation}
The interesting new phenomenon associated with the occurrence of a gradient
of the field strength along the field line ($L$) is the possibility to shape
different cooling trajectories according to 
\begin{equation}
\frac{d\varepsilon _{\parallel }}{\varepsilon _{\parallel }\left( 1/\tau \mp
1/\tau _{0}\right) }=2\frac{d\varepsilon _{\perp }}{\varepsilon _{\perp
}\left( 1/\tau \pm 2\varepsilon _{\perp }\varepsilon _{\parallel
0}/\varepsilon _{\perp 0}^{2}\tau _{0}\right) }\text{.}
\end{equation}

This new freedom opens the way to an optimization of the final stage of the
free energy extraction. The necessity of such an optimization is clearly
displayed by the analysis of the evolution of the anisotropy of the particle
energy distribution function. Consider a particle with initial parallel,
perpendicular, and total energies $\varepsilon _{||0}$, $\varepsilon _{\perp
0}$, and $\varepsilon _{0}$, respectively. Define the initial pitch angle $%
\vartheta $ by 
\begin{eqnarray}
\varepsilon _{||0} &=&\varepsilon _{0}\cos ^{2}\vartheta \text{,} \\
\varepsilon _{\perp 0} &=&\varepsilon _{0}\sin ^{2}\vartheta \text{.}
\end{eqnarray}
This gives $\varepsilon _{||0}+\varepsilon _{\perp 0}=\varepsilon _{0}$ by
construction. Equipartition of energy between the three directions of motion
corresponds to $\vartheta _{\text{EP}}\ =\arccos (\sqrt{1/3})$. For the
previous toroidal case we have found 
\begin{equation}
\frac{\varepsilon _{\perp }}{\varepsilon _{\parallel }}=\tan ^{2}\vartheta
\exp \left( \frac{t}{\tau }\right)
\end{equation}
This increase of the cyclotron energy at the expense of the parallel energy can
be controlled if we modulate the purely toroidal field and introduce a new
structural freedom with $L\left( s\right) $ continuously redistributing the
cyclotron energy into the parallel one during the slowing down process. This
study of the optimization of the magnetic and electric fields configuration
is left for a future work as it must be addressed after a careful assessment
of the limitations of the purely toroidal configuration. The next sections (V
and VI) are devoted to the analysis of these limitations which are of
dynamical and geometrical nature.

\section{Efficiency: geometrical limitations}

\begin{figure*}
	\includegraphics[width=\linewidth]{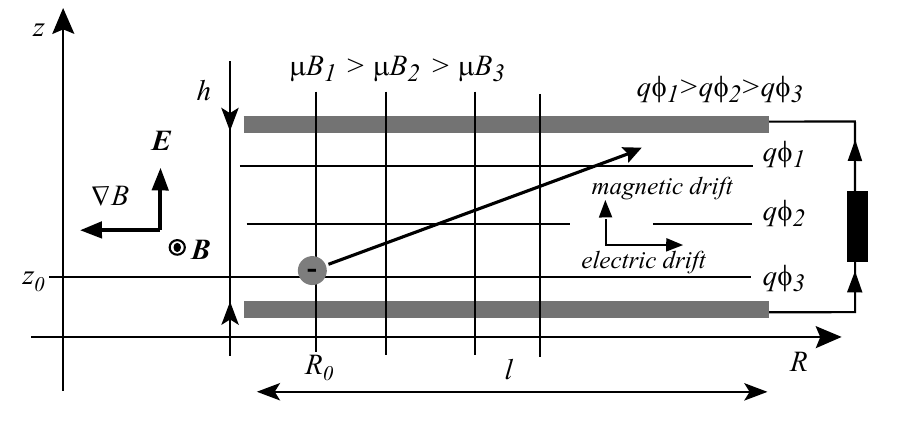}
	\caption{The trajectory of a particle as it moves between the electrodes includes both magnetic and electric drift motion.}
\end{figure*}

We can define the efficiency of free energy extraction as the fraction $\eta
\leq 1$ of extracted initial thermal energy

\begin{eqnarray}
\eta &=&1-\frac{\varepsilon _{\parallel }\left( t\right) +\varepsilon
_{\perp }\left( t\right) }{\varepsilon _{0}} \\
&=&1-\cos ^{2}\vartheta \exp \left( -\frac{2t}{\tau }\right) -\sin
^{2}\vartheta \exp \left( -\frac{t}{\tau }\right) \text{.}
\end{eqnarray}

This relation clearly displays the advantage of the possibility to control,
independently of the energy dynamics, the pitch angle dynamics $\vartheta$. 
Such a possibility is offered by a gradient of the field strength along the
field line ($L$) analyzed in Section IV. One major limit on the achievable
efficiency $\eta $ is the time available before a particle strikes a
boundary of the device. Following Figure~8, let $h$ be the width of the gap
between the two electrodes and $\ell$ be the radial extent of the circular
electrodes.

The transit from the inner edge $R=R_{0}$ to the outer edge $R=R_{0}+\ell $
takes time $t_{r}$ given by 
\begin{equation}
\frac{t_{r}}{\tau }=\log \left( 1+\frac{\ell}{R_{0}}\right) \text{.}
\end{equation}
A particle could also strike the axial boundary first. The time to transit
from $z=z_{0}$ to $z=z_{0}+h$ is

\begin{equation}
\frac{t_{z}}{\tau }=\log \frac{\sin ^{2}\vartheta +\sqrt{\sin ^{4}\vartheta
+4\cos ^{2}\vartheta (1-qEh/\varepsilon _{0})}}{2\left( 1-qEh/\varepsilon
_{0}\right) }
\end{equation}
where $Eh$ is the full voltage drop between the electrodes and $%
qEh/\varepsilon _{0}$ the ratio of that potential energy to the initial kinetic energy of the particle. Note that the square root is real when $\varepsilon
_{0}>qEh$. Then this first limit is determined by the efficiency $\eta $
that can be achieved in the lesser of $t_{r}$ and $t_{z}$ (or, in the case
where $t_{z}\notin \Bbb{R}$, it is determined by $t_{r}$ alone).

This can also be understood as a constraint on the system size required to
achieve a particular efficiency $\eta $. Consider, for example, the case in
which the radial size $\ell $ is limiting (rather than the axial size $h$).
If $\vartheta =\vartheta _{\text{EP}}$, $\ell $ and $\eta $ are related by 
\begin{equation}
\frac{\ell }{R_{0}}=\left( \sqrt{4-3\eta }-1\right) ^{-1}-1\text{.}
\end{equation}
Near-perfect efficiencies would require $\ell \gg R_{0}$. The axial gap $h$
must also be large enough in order to ensure a significant conversion
efficiency $\eta $, taking $h=z-z_{0}$ in Eq. (\ref{zz}) and a thermal
distribution with equipartition gives the relation 
\begin{equation}
\frac{qEh}{k_{B}T}=3/2-\left( \sqrt{4-3\eta }-1\right) ^{2}/2-\left( \sqrt{%
4-3\eta }-1\right)
\end{equation}
Apart from the geometric scaling, the other main constraint on this DEC
scheme has to do with the validity of the first order adiabatic drift theory
which is analyzed in the next section.

\section{Efficiency: dynamical limitations}

\begin{figure*}
	\includegraphics[width=\linewidth]{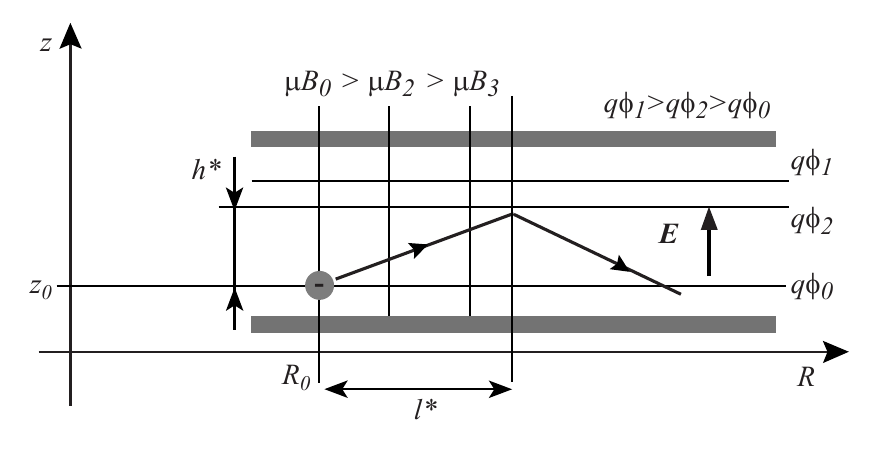}
	\caption{A dynamical limitation of this DEC scheme: if a particle is allowed to move too far without striking an electrode, the energy transfer will reverse and the particle will start to take energy from the fields rather than transferring energy to the fields.}
\end{figure*}

To identify the dynamical limit we consider the next order drift within the
framework of adiabatic theory: the second-order inertial drift $\mathbf{v}%
_{DI}$, which can be written as follows: 
\begin{equation}
\mathbf{v}_{DI}=\frac{\mathbf{B}}{qB^{2}}\times m\frac{d}{dt}\left( \frac{%
\mathbf{E}\times \mathbf{B}}{B^{2}}\right) \mathbf{=}\frac{m}{qB\tau }\frac{%
\mathbf{E}}{B}\text{.}
\end{equation}
Here $\tau $ is the same timescale introduced in Eq.~(\ref{to}). We
recognize here the usual polarization drift associated with the time
variation $\tau $. We see that this inertial drift is always along $q\mathbf{%
E}$ in the direction opposite to the magnetic drift and provides a limiting
effect to the previous first order conversion process. This second-order
drift does not affect the secular radial dynamics given in Eq.~(\ref{drif2}%
), but the secular axial dynamics in Eq.~(\ref{drif1}) becomes 
\begin{equation}
\frac{dz}{dt}=\left( \mathbf{v}_{D}+\mathbf{v}_{DI}\right) \cdot \mathbf{e}%
_{z}=\frac{2\varepsilon _{\parallel }+\varepsilon _{\perp }}{qE\tau }-\frac{%
mE}{qB^{2}\tau }\text{.} \label{eqn:dzdt}
\end{equation}
For any given particle, the transfer of energy from kinetic energy to the
electric field will reverse when $dz/dt=0$, at which point ($z-z_{0}=h^{*}$, 
$R=R_{0}+l^{*}$) the device will no longer operate as a DEC generator but as
an accelerator for that particle. This reversal of the energy transfer is
illustrated in Fig. 9.

The power transfer between the thermal energy and the electric energy is
given by the dot product $q\mathbf{E}\cdot \left( \mathbf{v}_{D}+\mathbf{v}%
_{DI}\right) $: 
\begin{equation}
qE\frac{dz}{dt}=\frac{2\varepsilon _{\parallel }+\varepsilon _{\perp }}{\tau 
}-\frac{mE^{2}/B^{2}}{\tau }\text{.}  \label{rev1}
\end{equation}
We define the drift energy as 
\begin{equation}
\varepsilon _{E/B}=\frac{m}{2}\frac{E^{2}}{B^{2}}\text{.}
\end{equation}
To study the limitation associated with this reversal of the energy
transfer, we now consider the case in which the $\mathbf{E}\times \mathbf{B}$
drift may contain a significant fraction of the kinetic energy, in which
case the leading-order expression for energy becomes 
\begin{equation}
\varepsilon \left( z,R\right) =\varepsilon _{\parallel }+\varepsilon _{\perp
}+\varepsilon _{E/B}+q\phi
\end{equation}
rather than Eq. (\ref{energy}). The leading-order expression for energy
conservation ought to be 
\begin{equation}
\frac{d}{dt}\left( \varepsilon _{\parallel }+\mu B+q\phi +\varepsilon
_{E/B}\right) =0  \label{rev}
\end{equation}
Thus Eqs. (\ref{ener2}) and (\ref{ener3}) are replaced by 
\begin{equation}
\frac{d\varepsilon _{\parallel }}{dt}=-\left[ \left( \mathbf{v}_{D}+\mathbf{v%
}_{DI}+v_{\parallel }\frac{\mathbf{B}}{B}\right) \cdot \mathbf{\nabla }%
\right] \left( q\phi +\mu B+\frac{m}{2}\frac{E^{2}}{B^{2}}\right)
\end{equation}
and 
\begin{equation}
\frac{d\varepsilon _{\perp }}{dt}=\left[ \left( \mathbf{v}_{D}+\mathbf{v}%
_{DI}+v_{\parallel }\frac{\mathbf{B}}{B}\right) \cdot \mathbf{\nabla }%
\right] \mu B
\end{equation}
completed by the inertial/polarization drift effect 
\begin{equation}
\frac{d\varepsilon _{E/B}}{dt}=\left[ \left( \mathbf{v}_{D}+\mathbf{v}%
_{DI}+v_{\parallel }\frac{\mathbf{B}}{B}\right) \cdot \mathbf{\nabla }%
\right] \left( \frac{m}{2}\frac{E^{2}}{B^{2}}\right)  \label{eonb}
\end{equation}
where 
\begin{equation}
\left( \mathbf{v}_{D}+\mathbf{v}_{DI}+v_{\parallel }\frac{\mathbf{B}}{B}%
\right) \cdot \mathbf{\nabla }=\left( \frac{2\varepsilon _{\parallel
}+\varepsilon _{\perp }}{qE\tau }-\frac{mE}{qB^{2}\tau }\right) \frac{%
\partial }{\partial z}+\frac{E}{B}\frac{\partial }{\partial R}\text{,}
\end{equation}
rather than Eq. (\ref{adia2}).

The evolution of the thermal parallel and perpendicular cyclotron energies
fulfill 
\begin{eqnarray}
\frac{d\varepsilon _{\parallel }}{dt} &=&-\frac{2\varepsilon _{\parallel }}{%
\tau }\text{,}  \label{rev2} \\
\frac{d\varepsilon _{\perp }}{dt} &=&\mathbf{-}\frac{\varepsilon _{\perp }}{%
\tau }\text{.}  \label{rev3}
\end{eqnarray}
Note that this $\varepsilon _{\perp }$ is the part of the kinetic energy in
the Larmor gyration, not the total kinetic energy in the perpendicular
direction (which also includes drift motion contribution $\varepsilon _{E/B}$%
). The evolution of $\varepsilon _{E/B}$ Eq. (\ref{eonb}) is given by 
\begin{equation}
\frac{d\varepsilon _{E/B}}{dt}=\frac{2\varepsilon _{E/B}}{\tau }\text{.}
\label{revn}
\end{equation}

It can be checked that Eqs. (\ref{rev},\ref{rev1}) are consistent with Eqs. (%
\ref{rev2},\ref{rev3},\ref{revn}) as $Ez=\phi $. These equations (\ref{rev2},%
\ref{rev3},\ref{revn}) can be integrated directly. The three components of
the kinetic energy evolves respectively according to 
\begin{eqnarray}
\varepsilon _{\parallel } &=&\varepsilon _{\parallel 0}\exp \left( -\frac{2t%
}{\tau }\right) \text{,} \\
\varepsilon _{\perp } &=&\varepsilon _{\perp 0}\exp \left( -\frac{t}{\tau }%
\right) \text{,} \\
\varepsilon _{E/B} &=&\frac{\varepsilon _{0}}{2C}\exp \left( \frac{2t}{\tau }%
\right) \text{,}
\end{eqnarray}
where we have defined the constant 
\begin{equation}
C\doteq \frac{\varepsilon _{0}B_{0}^{2}}{mE^{2}}\,\text{.}
\end{equation}
The transfer of kinetic to potential energy is 
\begin{eqnarray}
qE(z-z_0) &=&\varepsilon _{\parallel 0}\left[ 1-\exp \left( -\frac{2t}{%
\tau }\right) \right] \nonumber \\
&&+\varepsilon _{\perp 0}\left[ 1-\exp \left( -\frac{t}{\tau }\right)
\right] \nonumber \\
&&+\frac{mE^{2}}{2B_{0}^{2}}\left[ 1-\exp \left( \frac{2t}{\tau }%
\right) \right] .  
\end{eqnarray}
The key effect captured by the inclusion of $\mathbf{v}_{DI}$ is that the
increase in $\mathbf{E}\times \mathbf{B}$ flow energy, as the particle moves
outwards, tends to slow and eventually reverse the transfer from kinetic to
potential energy.

The time $t^{*}$ at which this reversal takes place is given by $dz/dt=0$ in
(\ref{rev1}) 
\begin{equation}
2\varepsilon _{\parallel }\left( t^{*}\right) +\varepsilon _{\perp }\left(
t^{*}\right) =\frac{mE^{2}}{B^{2}}=\frac{mE^{2}}{B_{0}^{2}}\exp \left( \frac{%
2t^{*}}{\tau }\right)
\end{equation}
so that we have to solve 
\begin{equation}
2C\cos ^{2}\vartheta +C\sin ^{2}\vartheta \exp \left( \frac{t^{*}}{\tau }%
\right) =\exp \left( \frac{4t^{*}}{\tau }\right) .  \label{cc}
\end{equation}
in order to find $t^*$. Numerical solutions of Eq.~(\ref{cc}) for the cases of $\vartheta = 0$, $\vartheta =\vartheta_{\text{EP}}$, and $\vartheta = \pi/2$ are shown in Fig.~10.
\begin{figure}
	\includegraphics[width=\linewidth]{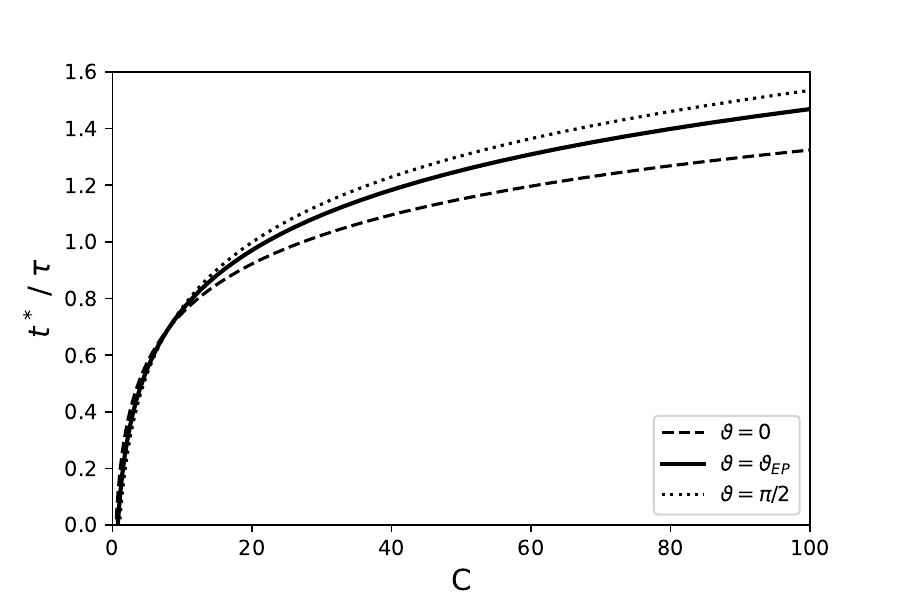}
	\caption{The critical time $t^*$ after which the energy transfer reverses, as a function of the dimensionless parameter $C$. The curves shown here take $\vartheta = 0$, $\vartheta = \vartheta_\text{EP}$, and $\vartheta = \pi/2$, respectively.} 
\end{figure}

The logarithmic behavior of this numerical solution describing the initial
equipartition case reflects the behavior of the particular solutions
associated respectively with $\vartheta =0$ and $\vartheta =\pi /2$:
\begin{eqnarray}
\left. \frac{t^{*}}{\tau }\right| _{\vartheta =0} &=&\log \left( 2C\right) /4%
\text{,} \\
\left. \frac{t^{*}}{\tau }\right| _{\vartheta =\pi /2} &=&\log \left(
C\right) /3\text{.}
\end{eqnarray}
The maximum free energy extraction efficiency in a large device, free of the
geometrical limitations analyzed in the previous section, is 
\begin{equation}
\eta ^{*}\left( E,B_{0},\varepsilon _{0},\vartheta \right) =1-\cos
^{2}\vartheta \exp \left( -\frac{2t^{*}}{\tau }\right) -\sin ^{2}\vartheta
\exp \left( -\frac{t^{*}}{\tau }\right)
\end{equation}
Numerical solutions for this maximum efficiency $\eta ^{*}\left( C\right) $ for a large device are shown on Fig.~11 for the cases of $\vartheta = 0$, $\vartheta =\vartheta _{\text{EP}}$, and $\vartheta = \pi/2$. 

\begin{figure}
	\includegraphics[width=\linewidth]{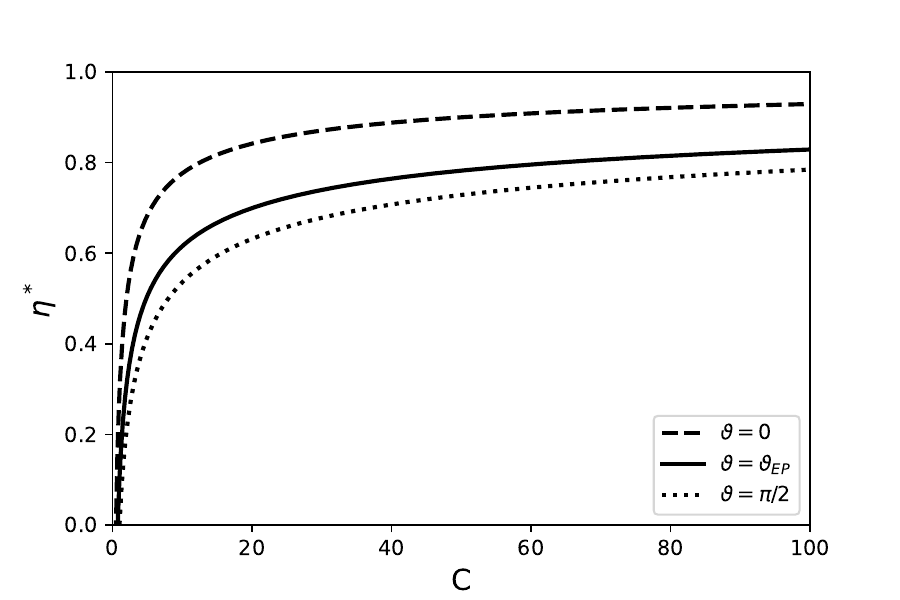}
	\caption{The maximum free energy extraction efficiency as a function of the dimensionless parameter $C$ for selected values of $\vartheta$. This maximum takes into account dynamical but not geometrical constraints (that is, it assumes an optimally large device). }
\end{figure}

The algebraic behavior of the type $1-\alpha C^{-\beta }$ of this numerical
solution describing the initial equipartition case reflects the behavior of
the particular solutions associated respectively with $\vartheta =0$ and $%
\vartheta =\pi /2$:
\begin{eqnarray}
\eta ^{*}\left( \vartheta =0\right) &=&1-C^{-\frac{1}{2}}/\sqrt{2} \\
\eta ^{*}\left( \vartheta =\pi /2\right) &=&1-C^{-\frac{1}{3}}
\end{eqnarray}
The efficiency of a real device will depend on the birth distribution of
charged particles both in $\left( \varepsilon _{0},\vartheta \right) $ space
and in spatial position $\left( z_{0},R_{0}\right) $. For any birth
distribution, it is clear that this effect will reduce the realizable
efficiency. This can be understood as a limitation on the current-voltage ($%
I $-$U$) curve describing the operation of an adiabatic DEC generator. The
total voltage drop of the configuration can be increased either by
increasing $E$ or by increasing $h$. For any given starting condition $%
\left( \varepsilon _{0},\vartheta \right) $, if either $E$ or $h$ is
increased beyond some threshold, the particle will not reach the collection
plate before its trajectory reverses. Note, however, that although
increasing either $E$ or $h$ will increase the total voltage drop, and
either will eventually cause electrons to turn before they reach the
negative electrode, these two parameters influence the dynamics in different
ways. $E$ and $h$ both change the total energy that must be extracted before
a particle can traverse a given axial distance, but changing $E$ also
modifies the feedback from the inertial drift.

If all electrons reach the negatively charged electrode, then the total
device current $I$ is set by the rate of ionization events or the rate of
charged particles incoming flow, and $I$ is independent of $U$. However, as $%
U$ increases, there is a threshold (which will depend on the birth
distribution of the charged particles and of the size, field strength and
shape of the configuration) where $I$ will quickly drop off as a result of
electrons turning before they can reach the negative electrode.
Qualitatively, this will produce an $I$-$U$ characteristic like the one
pictured in Figure~12, though one should keep in mind that the details of
this curve will depend on not only the details of the birth distribution but
also on what is held fixed when the voltage is increased. For any particular
case, this curve makes it possible to determine the highest-power operating
point; one can draw the set of isopower hyperbolae $IU=\text{constant}$ and
select the one that is tangent to the convex $I$-$U$ generator
characteristic.

\begin{figure}
	\includegraphics[width=\linewidth]{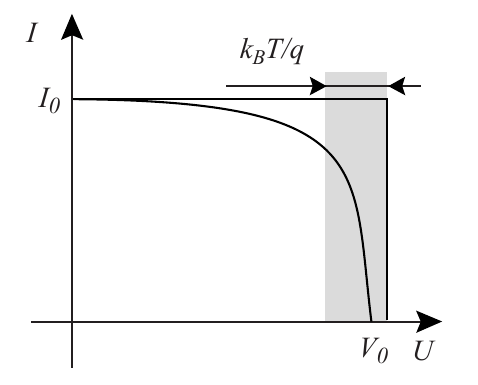}
	\caption{The current-voltage characteristic for a DEC generator of this type is initially flat, with little dependence of the current on the voltage. At higher voltages, the output current drops as a function of voltage.} 
\end{figure}

In the most idealized case, in which the birth distribution is monoenergetic
and all particles are born at the same spot and with the same $\vartheta$,
the $I$-$U$ curve is a step function with constant $I=I_{0}$ for all $%
U<V_{0} $, then $I=0$ for all $U\geq V_{0}$ (with $V_{0}$ corresponding to
the threshold at which the particles turn before reaching the boundary $%
dz/dt=0$). 
In this case, the highest-power operating point corresponds to
the corner of the $I$-$U$ curve, with $P=I_{0}\ V_{0}$. For the case of a
radiation driven plasma generation, this power would be a linear function of
the absorbed radiation (since this would determine the total number of
ionization events).
From Eq.~(\ref{eqn:dzdt}), $dz/dt = 0$ implies that $m E^2 / B^2 = \mathcal{O} (k_B T)$, so that $V_0 = E h = \mathcal{O} (B_0 v_{th} h)$, where $v_{th}$ is the thermal velocity. 

\begin{figure}
	\includegraphics[width=\linewidth]{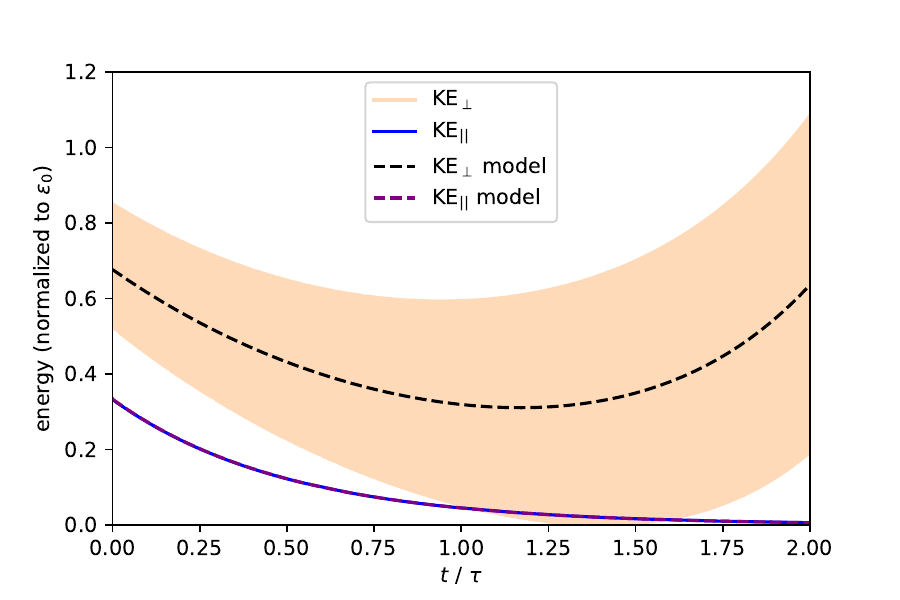}
	\caption{Numerical validation of the analytic model using a single-particle simulation. This case had $C = 50$, from which we would expect a turning point at $t \approx 1.253 \tau$. The region colored in by the $\text{KE}_\perp$ curve is the result of fast oscillations due to the gyrophase-dependent orientations of the Larmor motion and the drift motion. }
	\label{fig:numericalValidation}
\end{figure}

The previous analytic calculations can be validated numerically using
single-particle simulations. Figure~\ref{fig:numericalValidation} shows one such example.

The simulation showed in the figure used a single-particle Boris pusher and
had $C=50$ and $\vartheta =\vartheta _{\text{EP}}$. Note that for these
parameters, Eq.~(\ref{cc}) can be solved numerically to yield $t^{*}\approx
1.253\tau $. This is consistent with the turning point seen in the
simulation. The fluctuations in the numerically observed KE$_{\perp }$ about
the model are the result of the gyrations of the particle. Depending on the
gyrophase, the Larmor motion can have either positive or negative radial
components.

The energy in the perpendicular motion can only be decomposed into gyration
energy and drift-motion energy on average. The kinetic energy used for the
model in the figure includes both $\varepsilon _{\perp }$ and $mE^{2}/2B^{2}$%
.

These simulations confirm the validity of the previous analytical model and
points toward the necessity to consider an additional possibility to convert
the cyclotron energy into the parallel energy. We have seen in Section IV
that a gradient of the field strength along the field line offers such a
possibility and this optimization strategy will be explored in a forthcoming
study.

\section{Radiation flux and plasma flow heat conversion}

In this section we briefly describe two types of adiabatic DEC
implementation in $^{2}$D/$^{3}$T tokamak reactors and in aneutronic $^{1}$%
P/$^{11}$B reactors.

\begin{figure*}
	\includegraphics[width=\linewidth]{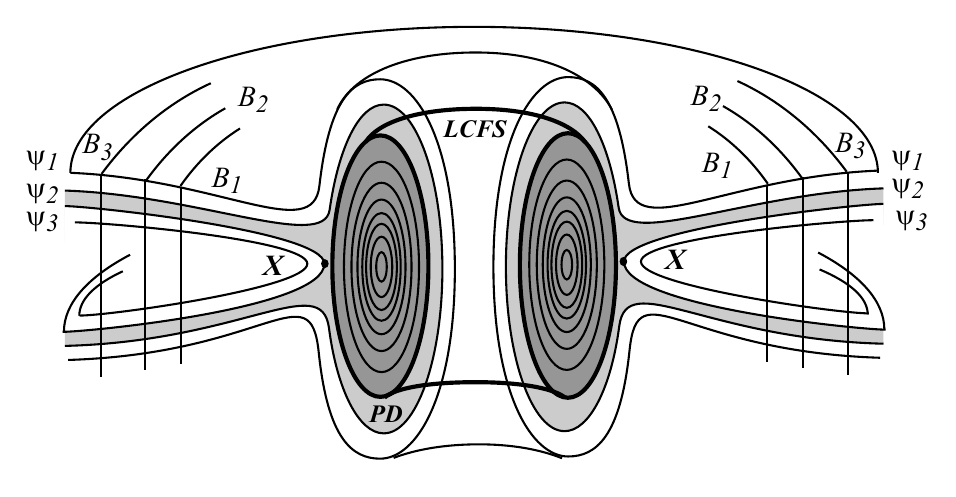}
	\caption{A schematic of the magnetic flux surfaces in a tokamak, with the last closed flux surface (LCFS) marked.} 
\end{figure*}

In a tokamak reactor the hot thermonuclear plasma diffuses from the confining
closed field line to the open field lines of the divertor. The last closed
flux surface (LCFS) defines the boundary of the confining toroidal
configuration. The set of diverted field lines defines open magnetic surfaces
illustrated in Fig.~14.

Far from the X point line, the magnetic field in the outer part of the
divertor, illustrated in Fig. 14, is toroidal and a dedicated set of upper
and lower electrodes can polarize these magnetic surfaces such that they
become also equipotential surfaces $\Psi_{1}$, $\Psi_{2}$, $\Psi_{3}$. In
doing so we have realized the geometry of a toroidal generator and we can
envision to extract part of the free enthalpy of the divertor plasma flow in
such a configuration. This geometry of an advanced divertor requires a far
more detailed analysis which is left to a future work.

Let us now consider the case of rotating mirror $^{1}$P/$^{11}$B reactors.
Because of the high temperature, the thermonuclear plasma column is strongly
radiating both in the microwave range, as a result of electron
cyclotron/synchrotron emission, and in the UV-X range, as a result of
electron bremsstrahlung. Keeping the axisymmetric geometry (Fig. 6-(b)) we
can design a conversion blanket all around this magnetized radiating plasma
column where the escaping intense radiation will ionize a vapor, for example
cesium vapor, and heat the associated plasma. The escaping radiation can
also be absorbed by metal target plates, for example tungsten plates, and
the resulting hot metal plate can act as an hot electron source whose energy
is converted by the E cross B poloidal configuration of the conversion
blanket.

Various toroidal, poloidal and mixed declinations of the E cross B
configurations can be adapted to both $^{2}$D/$^{3}$T tokamak reactors and
in aneutronic $^{1}$P/$^{11}$B reactors. The identification of the most
relevant designs can not be addressed on the basis of the simple discussion
of this section and is left for a future study.

\section{Conclusion}

To summarize our results, we suggests a novel technique for converting power
from plasma and radiation to electricity. A technology that could
efficiently capture and convert power from plasma and radiation to
electricity could be useful in a wide variety of applications. For example,
magnetic confinement fusion typically involves heating plasmas to
temperatures at which radiative losses can be significant. The ability to
recapture this energy could be most critically important for reactors
burning aneutronic fuels, which often require higher temperatures with
correspondingly higher radiative losses. For instance, bremsstrahlung and
synchrotron losses are a major hurdle for economical p-$^{11}$B fusion \cite{Putvinski2019, Kolmes2022PowerBalance, Ochs2022, Ochs2024PowerBalance}.
In very hot fusion plasmas, measures have been proposed to suppress this radiation primarily through plasma absorption and redirection to kinetic energy, as well as through suppression of the population of high-energy electrons \cite{Volosov2006, Mlodik2023, Munirov2023, Ochs2024TailSuppression}. However, the direct energy conversion schemes here could, in principle, act synergistically with these other techniques. 
The manipulation of the plasma 6D phase space has been recognized as necessary for making economical fusion through high temperature aneutronic fusion approaches in general \cite{Qin2024}, and the techniques proposed here can be imagined working either separately or in concert with these phase space techniques.

For a hot plasma such that $k_{B}T\sim \varepsilon _{0}$, the design of a
generator implies the choice of (\textit{i}) the geometrical characteristics $%
\left( R_{0},l,h\right) $, and of (\textit{ii}) the field strengths $\left(
E,B_{0}\right) $. We have found that the various physical and geometrical
constraints are in fact functions of the two control parameters $\tau
=B_{0}R_{0}/E$ and $C=\varepsilon _{0}B_{0}^{2}/mE^{2}$. The description of
the poloidal-geometry dynamics requires a third control parameter $\tau
_{0}=L\sqrt{2m\varepsilon _{\parallel 0}}/\varepsilon _{\perp 0}$. The
poloidal geometry displays an additional advantage over the toroidal one as
the cyclotron energy can be continuously converted into parallel energy.
Such a possibility is advantageous because E cross B configurations
convert the parallel energy to DC electric power at a far faster rate than
the cyclotron (perpendicular) energy. The design of an optimal field is clearly the next question to be addressed to identify efficient E cross B conversion schemes for both $^{2}$D/$^{3}$T tokamak and $^{1}$P/$^{11}$B advanced reactors.

The idea of poloidal and toroidal DEC rests on the tendency of the $\nabla B$
drift to carry particles in different directions depending on the sign of
their charge. Its implementation in $^{1}$P/$^{11}$B advanced reactors relies
on radiative ionization and heating. Radiation can deposit significant
energy, after ionization, in the resulting ions and electrons. 

Note that the transfers of energy considered here -- e.g., between kinetic energy and electric fields -- are all adiabatic. 
This can be contrasted with work (such as the alpha-channeling concept \cite{Fisch1992}) that instead accomplishes this transfer using resonant interactions. 

This adiabatic transfer process -- like all energy-conversion processes -- will not be perfectly efficient. In this paper, we have considered two constraints on the performance of an adiabatic DEC device. The first is a matter of geometry: the charged particles need enough space to give up as much kinetic energy as possible before hitting a boundary of the device. Even in the absence of any other constraints, this is enough to prevent any realistic device from attaining perfect efficiency, since progressively better efficiencies require exponentially increasing device sizes.

We also considered a dynamical constraint, wherein inertial drifts
eventually slow (and subsequently reverse) the energy extraction process.
The characteristic timescale involved depends on the parameter $C$. The
highest efficiencies require higher values of $C$.

There are a variety of additional constraints and engineering challenges not
considered here. For example, this kind of device relies on the presence of
a large enough population of neutral particles to absorb and be ionized by
the incoming radiation. However, if that population were too large,
collisions between charged and neutral particles would degrade the
efficiency of the device. 

In addition, a birth distribution of charged particles that is anything
other than a delta function in $r$, $z$, $\varepsilon_0$, and $\vartheta$
will make it more difficult to efficiently tune the device parameters. For
example, this would mean that different particles' trajectories would turn
at different axial and radial positions.

However, these engineering challenges can likely be mitigated by careful
control over the device geometry and the composition of the neutral
population. 
The refinement and optimization of these ideas are planned in future work. 
Moreover, even relatively modest efficiencies could be an exciting development, particularly for applications in which the relevant radiation has wavelengths not amenable to other conversion techniques.

\begin{acknowledgments}
This work was supported by ARPA-E grant no. DE-AR0001554. 
JMR further acknowledges the hospitality of Princeton University and support from the Andlinger Center for Energy and the Environment through an ACEE fellowship, under which this work was initiated.
EJK acknowledges the support of the DOE Fusion Energy Sciences Postdoctoral Research Program, administered by the Oak Ridge Institute for Science and Education (ORISE) and managed by Oak Ridge Associated Universities (ORAU) under DOE contract no. DE-SC0014664. 

\end{acknowledgments}


\bibliography{../DECRefs.bib}
\end{document}